\documentclass[12pt,preprint]{aastex}

\newcommand{\beq}{\begin{equation}}
\newcommand{\eeq}{\end{equation}}
\newcommand{\bdm}{\begin{displaymath}}
\newcommand{\edm}{\end{displaymath}}

\slugcomment{To appear in the Astrophysical Journal
  (Letters) \\ {\it Received 3 February 2004; accepted 1 March 2004}}

\begin{document}

\title{The Coronal X-ray Spectrum of the Multiple Weak-Lined T Tauri
  Star System HD 98800} 

\author{Joel H. Kastner\altaffilmark{1}, David
  P. Huenemoerder\altaffilmark{2}, Norbert S. Schulz\altaffilmark{2}, 
  Claude R. Canizares\altaffilmark{2}, Jingqiang
  Li\altaffilmark{1} and David A. Weintraub\altaffilmark{3} 
}

\altaffiltext{1}{Chester F. Carlson Center for
Imaging Science, Rochester Institute of Technology,
Rochester, NY 14623; jhk@cis.rit.edu}
\altaffiltext{2}{MIT Center for Space Research, 70 Vassar
  St., Cambridge, MA, 02139}   
\altaffiltext{3}{Dept.\ of Physics and Astronomy, Vanderbilt University,
Nashville, TN, 37235}

\begin{abstract}

We present high-resolution X-ray spectra of the multiple
(hierarchical quadruple) weak-lined T Tauri star system HD
98800, obtained with the High Energy Transmission Gratings
Spectrograph (HETGS) aboard the Chandra X-ray Observatory
(CXO). In the zeroth-order CXO/HETGS X-ray image, both
principle binary components of HD 98800 (A and B, separation
$0.8''$) are detected; component A was observed to flare during the
observation. The infrared excess (dust disk)
component, HD 98800B, is a factor $\sim4$ fainter in X-rays
than the apparently ``diskless'' HD 98800A, in quiescence. 
The line ratios of He-like species (e.g., Ne
{\sc ix}, O {\sc vii}) in the HD 98800A spectrum indicate
that the X-ray-emitting plasma around HD 98800 is in a
typical coronal density regime ($\log{n} \stackrel{<}{\sim} 11$). We
conclude that the dominant X-ray-emitting component(s) of HD 
98800 is (are) coronally active. The sharp spectral
differences between HD 98800 and the classical T Tauri star
TW Hya demonstrate the potential utility of high-resolution X-ray
spectroscopy in providing diagnostics of pre-main
sequence accretion processes. 
 
\end{abstract}


\section{Introduction}

The Einstein and ROSAT missions established the ubiquity of
X-ray emission from low-mass, pre-main sequence (PMS)
stars. These early X-ray observations, and subsequent
observations by the ASCA satellite X-ray observatory, left
unsolved the fundamental problem of the physical origin of the
X-ray emission, which could be solar-type coronal activity, star-disk
interactions, or some combination of these mechanisms.
Much of the X-ray data pointed to the likely importance of
magnetic activity (Feigelson \& Montmerle 1999).  To
make further progress, it is necessary to determine the temperature
distributions, densities, and elemental abundances of the
X-ray-emitting plasmas of PMS stars, so as to provide
constraints on models of X-ray emission from coronal and
star-disk interactions and to compare with, e.g., the physical
conditions characterizing well-established stellar coronal
X-ray sources (such as RS CVn systems and other close binaries).

With the advent of X-ray gratings spectrometers aboard the
Chandra X-ray Observatory and the XMM-Newton satellite
observatory, astronomers are now beginning to explore the
X-ray spectral characteristics of PMS stars. In this
regard, the TW Hya
Association (TWA) represents an especially useful young PMS
cluster. The TWA is a group of about 30 PMS
stars  (Zuckerman et al.\ 2001 and
references therein) located only $\sim50$ pc from the Sun and far from the
nearest star forming clouds (Kastner et al.\
1997). Its age ($\sim5-10$ Myr; e.g., Weintraub et al.\
2000, Zuckerman et al.\ 2001) likely 
corresponds to the epoch of Jovian planet formation in the
early solar system.
The TWA's proximity, relatively large ratio of X-ray to
bolometric luminosity ($\log{L_X/L_\star} \sim 3$, where
$L_X$ is measured within the ROSAT spectral bandpass of 0.1
to 2.4 keV; Kastner
et al.\ 1997), and lack of cloud absorption lead to 
uniformly high X-ray fluxes among its member stars. In terms of
ROSAT X-ray spectral properties, the TWA appears to represent a
transition stage between cloud-embedded PMS stars and the
zero-age main sequence (Kastner et al.\ 2003). 

The CXO/HETGS spectrum of TW
Hya itself is perhaps the most extreme
and intriguing of the many Chandra/HETGS spectra of X-ray
active stars obtained to date, in several key respects
(Kastner et al.\ 2002): (1) Ne is highly overabundant and Fe
severely underabundant, even in comparison with stars
exhibiting strong coronal abundance anomalies; (2) the
temperature distribution derived from fluxes of
temperature-sensitive emission lines 
is sharply peaked, at $\log T = 6.5$; (3) perhaps
most significantly, density-sensitive line ratios of Ne~{\sc
ix} and O~{\sc vii} indicate plasma densities $\log n
\sim 13$. This is more than an order of
magnitude larger than density estimates similarly obtained
for coronally active late-type stars. 

TW Hya remains the TWA's only unambiguous example of an
actively accreting (i.e., ``classical'') T Tauri star. Given
the evidence that TW~Hya likely is surrounded by a
circumstellar disk from which it is still accreting
(e.g., Muzzerole et al.\ 2000), we used the Chandra/HETGS results
to explore the hypothesis that X-ray emission from classical
T Tauri stars might originate from accretion streams that
connect the circumstellar disk to the star (Kastner et al.\
2002). Both the density range and the characteristic
temperature of X-ray emission obtained from modeling the
CXO/HETGS spectrum are consistent with recent models of
magnetospheric accretion onto T Tauri stars (Kuker,
Henning, \& Rudiger 2003).

In this {\it Letter}, we report on the results of CXO/HETGS
observations of a second well-studied TWA member, HD~98800
(TV Crt). HD 98800 is a hierarchical quadruple, weak-lined T
Tauri Star (wTTS) and is one of the best examples of
a solar-mass ``Vega-type'' (infrared excess) system
(Zuckerman \& Becklin 1993; Sylvester et al.\ 1996; Soderblom et
al.\ 1996, 1998). HD 98800 is a visual double with a separation of
$\sim0.8''$. Each visual component is itself
a spectroscopic binary, and one of these (HD 98800B) is double-lined
(Soderblom et al.\ 1996). Although the HD 98800B binary system
harbors a dust disk, the system is apparently
non-accreting; models indicate that the dust disk has an
inner gap extending to $\sim$2 AU (Prato et al.\ 2001). The
HD 98800 system displays an X-ray luminosity similar to that of
its fellow TWA member TW~Hya (Kastner et al.\ 1997).
Given that HD 98800 does not appear to be accreting, it
makes an excellent target for further 
investigation into the origins of X-ray emission from PMS stars
with CXO/HETGS. 

\section{Observations}

We observed HD 98800 with Chandra/HETGS for 98.9 ks on 2003 March 7
(observation identifier 3728) in the default configuration (timed
exposure, ACIS-S detector array) and under nominal operating
conditions.  Data were re-processed with Chandra Interactive Analysis
of Observations (CIAO; version 3) software to apply updated
calibrations, and events were cleaned of the detector artifacts on CCD
8 (``streaks'').  We applied subpixel event position corrections to
the zeroth-order events, following an algorithm for CXO
back-illuminated CCD data described in Li et al.\ (2003). Spectral
responses were generated with CIAO; corrections were also made for
ACIS contamination.  Lines were measured with ISIS\footnote{ISIS is
available from {\tt http://space.mit.edu/CXC/ISIS}} (Houck \& DeNicola
2000) by convolving Gaussian profiles with the instrumental response,
and emission measure and abundances were modeled with custom
ISIS programs (see, e.g., Huenemoerder et al.\ 2003b, for
a detailed description of the technique).

The resulting Chandra/HETGS spectral image of the HD 98800A/B binary
yielded 5163 zeroth order counts, 4099 first order medium energy
grating (MEG) counts, and 1336 first order high energy grating (HEG)
counts.  The integrated flux (1.5-25 \AA) obtained from the first-order
MEG and HEG data is
$2.5\times10^{-12}\,\mathrm{ergs\,cm^{-2}\,s^{-1}}$
($1.6\times10^{-3}\,\mathrm{photons\,cm^{-2}\,s^{-1}}$), corresponding
to a luminosity $L_x=6.9\times10^{29}\,\mathrm{ergs\,s^{-1}}$ at the
distance to HD 98800 ($D=48$ pc; Favata et al.\ 1998).

\section{Results}

\subsection{Zeroth-order image and light curve}

Zeroth-order CXO/HETGS images of HD 98800, before and after
applying subpixel event position corrections, are presented
in Fig.\ 1, alongside Keck Telescope images obtained in the
thermal infrared (Prato et al.\ 2001). The primary
components (HD 98800A and B) are well resolved in the
zeroth-order image, following event position
correction. In addition, light curves of the zeroth order sources and
nearby background demonstrate that component A flared during the
course of the observation (Fig.\ 2), while component B did not display
measurable variations in count rate. Prior to the onset of the flare, 
the CXO/HETGS count rate of HD 
98800A was a factor $\sim4$ larger than that of HD 98800B. Fig.\ 2
demonstrates that the HD 98800A flare was seen predominantly in
hard ($<7$ \AA) X-rays. 

\subsection{First-order (MEG+HEG) spectrum}

In Fig.\ 3 we present the spectrum of HD 98800A+B\footnote{Our initial
tests of a modified event repositioning technique --- in
which different position corrections are used for
front-illuminated and back-illuminated devices (Li et al.\
2004) --- were inconclusive as to potential
improvements in the spatial or spectral resolution of the dispersed
spectra. Hence, we present here the combined dispersed
spectrum of components A and B.} from 2
\AA\ to 25 \AA.  In the range from 12 \AA\ to 25 \AA, the
spectrum is dominated by emission lines from highly ionized
Ne, O, and Fe. Lines of O {\sc
viii} (16.0, 19.0 \AA), O {\sc vii} (21.6, 22.0 \AA), Ne {\sc x}
(12.1 \AA), Ne {\sc ix} (13.4, 13.7 \AA), and Fe {\sc xvii}
(15.0, 17.1 \AA) are particularly prominent. Several weaker
lines (e.g., Ne {\sc ix}, Ne {\sc x}, Mg {\sc xi}, Mg {\sc
xii}, Si {\sc xiii}, Ar {\sc xvii}) are clearly detected
shortward of 12 \AA, as well. 

In Fig.\ 4, we display narrow spectral regions around the
13.4, 13.55, and 13.7 \AA\ triplet (forbidden,
intercombination, and resonance lines, respectively;
hereafter {\it f, i, r}) of the He-like ion Ne {\sc ix}. In
this triplet, as in the He-like O {\sc vii} and Mg {\sc xi}
triplets (not shown), the intercombination line is the
weakest of the three lines, and the {\it f:i} ratio lies
between $\sim0.5$ and $\sim1.0$ (see also Huenemoerder et
al.\ 2003a). For the Ne {\sc ix} triplet, the ({\it f:i})
ratio is diagnostic of electron density over the range
$\log{n} \sim 11-13$ and, in the case of HD 98800, the Ne {\sc
  ix} {\it f:i} ratio suggests a density at the lower end of
this range. 

To better constrain the possible range of $n$
for HD 98800 we used
models, based on the Astrophysical Plasma Emission Database
(APED; Smith et al.\ 2001), in which the He-like triplet line
emissivities are calculated as functions of density and
temperature (Brickhouse, pvt.\ communication). For HD 98800,
the comparison between the measured Ne {\sc ix} {\it f:i} ratio 
and the APED-based model calculations indicates
$\log{n} \sim 11.25$, with a rather firm upper
limit of $\log{n} < 12$ (Fig.\ 5). From the O {\sc vii} {\it f:i}
ratio, which has a useful diagnostic 
range of about $\log{n} \sim 9.5-12$, we derive an upper
limit of $\log{n} < 11.5$. (These upper limits need not be
identical, since the lines are formed at different
temperatures, and may not be spatially coincident.)

From emission measure and abundance modeling of HD 98800, we
find O, Ne, and Fe abundances (relative to solar) of
$\sim0.3$, $\sim1.0$, and $\sim0.2$.  Fig.\ 4 serves as a
qualitative comparison of the relative abundances of Ne and
Fe for HD 98800 and TW Hya. Specifically, since Ne {\sc IX}
and Fe {\sc XVII} form at similar temperatures, their ratio
is primarily sensitive to relative abundance; Fig.\ 4 thus
illustrates that the relative overabundance of Ne with
respect to Fe is not as extreme for HD 98800 as for TW Hya.

\section{Discussion}

It is intriguing that the apparently ``diskless'' component
of the HD 98800 system, HD 98800A, appears to be the
stronger X-ray source (Fig.\ 1). Given the relatively small visual
extinction toward the 
system ($E(B-V) = 0.10$; Sylvester et al.\ 1996)
and the inference that the HD 98800B binary is 
viewed at intermediate inclination angle (Prato et al.\
2001), it seems unlikely that this difference in apparent
X-ray luminosities is due to differential
intervening absorption. Instead, it appears that A is
intrinsically brighter in X-rays than B, with a difference
of $\Delta (\log{L_X/L_{\rm bol}}) \approx 0.5$ between
components (see \S\S 2 \& 3.1 and Prato et al.\ 2001). While this result
appears consistent with a recent study of Orion showing that
disk-enshrouded pre-main sequence (PMS) stars are, in general, weaker
X-ray sources than ``diskless'' PMS stars (Flaccomio et al.\
2003), one must first
demonstrate that the quiescent emission from both components
does not show long-term variability, and that other binary TTS 
resolvable by Chandra show X-ray flux ratios that are
likewise anticorrelated with their relative IR excesses.

In terms of the {\it fir} line ratios
of He-like ions, and its Ne to Fe line ratios, the CXO/HETG X-ray
spectrum of HD 98800 resembles 
those of ``classical'' coronal sources, such as II Peg, UX
Ari, and HR 1099 (see, e.g., Fig.\ 6
in Kastner et al.\ 2002). Hence --- although it remains to
determine {\it fir} line ratios 
for additional weak-lined T Tauri stars --- {\it the {\rm fir} line
ratio results for HD 98800 provide some of the strongest
evidence to date that the X-ray emission from such stars is
coronal in origin.}

There is a marked contrast between the {\it fir} ratios in
the X-ray spectra of HD 98800 and TW Hya, however (Fig.\
4). For the latter (classical T Tauri) star, the forbidden
line is by far the weakest of the O {\sc vii} and Ne {\sc
ix} triplets, and the {\it i:r} ratio is near unity in each
case (the Mg {\sc xi} triplet is anomalously weak in the TW
Hya spectrum; Kastner et al.\ 2002). Fig.\ 5 shows that,
assuming the UV radiation field incident on the X-ray
emitting plasma of TW Hya is not strong (see below), the
plausible density regimes of Ne {\sc ix} line formation are
non-overlapping, for TW Hya and HD 98800; i.e., the Ne {\sc ix}
{\it f:i} ratio in the spectrum of TW Hya requires $\log{n}
> 12$.

Furthermore, whereas the differential emission measure (DEM) distribution
of TW Hya is sharply peaked at $\log{T} \sim 6.5$ (Kastner
et al.\ 2002), the DEM distribution of HD 98800 is
relatively flat over the temperature range $\log{T} \sim
6.4$ to $\log{T} \sim 7.0$. The latter behavior is more like
that of coronal sources (Huenemoerder et
al.\ 2003a,b) although --- unlike such sources --- HD 98800
evidently lacks strong emissivity around $\log{T}=7.2-7.6$.
This may be an indication of the evolutionary status of the
corona, in that the dynamo is not yet as strong as in the
coronally active binaries, or that the flare frequency,
which seems to drive the hotter peak, is not as high.

The overall X-ray spectral similarity between HD 98800 and coronally
active stars makes the CXO/HETG X-ray spectrum of the
classical TTS TW Hya that much more remarkable.
As argued by Kastner et al.\ (2002), both the plasma
densities implied by line ratios of He-like ions and its
sharply peaked (and rather low) X-ray temperature
distribution point to accretion as a likely source of some or all of its
X-ray emission. The modeling of Kuker et al.\ (2003) lends
additional credence to this argument. As TW Hya evidently is
actively accreting (based on its strong H$\alpha$
emission and its UV and near-infrared excesses) whereas
HD 98800 is not (based on these same accretion indicators),
the sharp distinction between their X-ray spectral
characteristics appears to further support the hypothesis
that the X-ray emission from TW Hya --- and, by extension,
other classical TTS --- may be generated, at least in part,
via accretion.

On the other hand, these two TWA stars are similar in
X-rays, in certain respects. For example, TW Hya was also observed to
flare, with flare characteristics (e.g., rise time, decay
time, peak flare to quiescent count rates) similar to
those of HD 98800 (Kastner et al.\ 2002). In addition,
although TW Hya is by far 
the most extreme star thus measured by HETG in terms of its
Ne/Fe abundance ratio, the X-ray spectrum of HD 98800 shows
similar abundance patterns overall (Huenemoerder et al.\ 2003b). 

These similarities would appear to cast some doubt on the
accretion hypothesis for TW Hya.  It is possible, for
example, that UV radiation, generated in accretion streams
onto TW Hya, depletes
the populations of atomic levels responsible for
the forbidden line component of the He-like triplets,
thereby driving the line ratios to their high-density limits (e.g.,
Ness et al.\ 2002). Although this phenomenon is reasonably
well established in the case of the intense UV fields of
X-ray-luminous O stars, it is less clear that it is a viable model
in the case of accreting cTTS. The relatively weak UV fields
of such stars likely would require that the X-ray-emitting plasma
be in very close proximity to the UV source -- effectively
placing the point of X-ray generation within (or very near)
the accretion stream itself. 

Whatever their origin, the contrasting results for the
density-sensitive line ratios of He-like ions in the X-ray
spectra of TW Hya and HD 98800 (Fig.\ 5) suggest that
fundamentally different physical conditions characterize the
X-ray emitting plasmas of cTTS and wTTS. 
CXO/HETG observations of additional cTTS and wTTS systems, as well
as detailed physical models of UV-irradiated coronal
plasmas, are now required to establish whether these He-like
triplet line ratios are probing X-ray emitting plasma in
accretion funnels or are, instead, diagnostic of the
intensity of accretion-powered UV and its proximity to the
corona of the accreting star.

\acknowledgements{Support for this research was provided by 
contracts SV3--73016 (CXC) and NAS8--01129 (HETG) to MIT.}

\clearpage

\begin{figure}[htb]
\includegraphics[scale=1.0,angle=0]{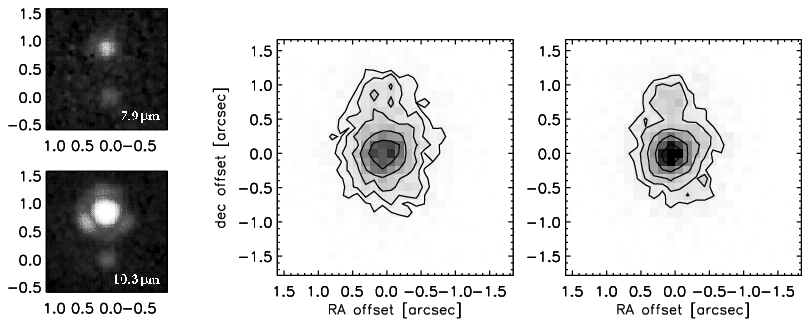}
\caption
{Left panels:
  Keck Telescope mid-infrared images of HD 98800 (Prato et al.\
  2001). Center and right
  panels: Chandra/HETGS zeroth-order X-ray images of HD 98800,
  before (center) and after (right) application of subpixel
  event relocation. In each X-ray image, contour levels 
  are 0.05, 0.1, 0.2, 0.4, 0.7 of the peak. The
  pixel size in these images is 0.125$''$. 
  HD 98800A, the stronger X-ray source but
  weaker mid-IR source, lies at offsets (0,0) in all four
  panels.} 
\end{figure}

\begin{figure}[htb]
\includegraphics[scale=0.6,angle=0]{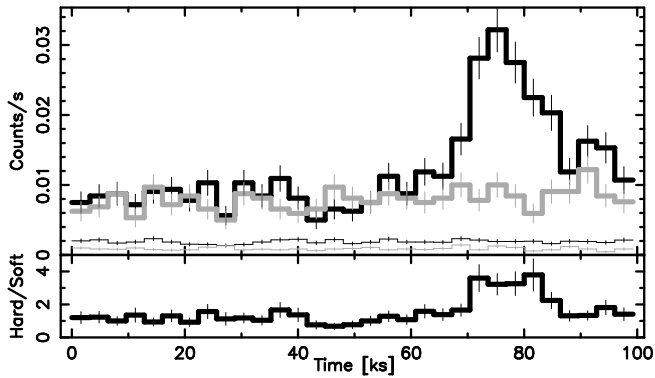}
\caption{Top panel: CXO/HETGS light curves of HD 98800 in
  ``hard'' (1.7--7.0 \AA; black) and ``soft'' (15.0--25.0
  \AA; grey) bands, for a bin size of 3200 s. Counts in
  dispersed spectral orders 1--3 were combined to 
  generate these plots. The thin black and grey curves are
  the background count rates in the hard and soft bands,
  respectively. Bottom panel: the ratio of hard to soft
  count rates.}
\end{figure}

\begin{figure}[htb]
\includegraphics[scale=0.75,angle=0]{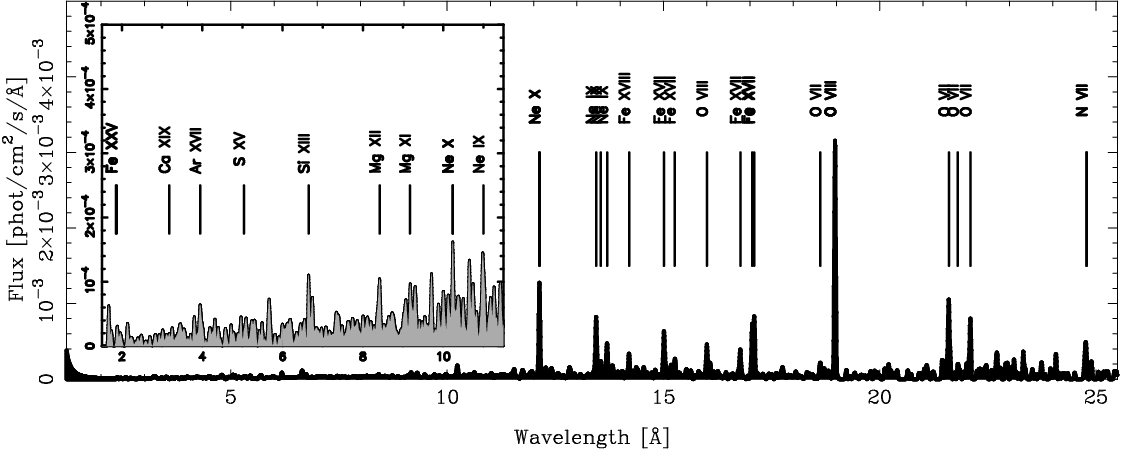}
\caption
{The combined HEG+MEG first-order spectrum of HD 98800. The
  inset shows the region from $\sim1.5$ \AA\ to $\sim12$ \AA.}
\end{figure}

\begin{figure}[htb]
\includegraphics[scale=.7,angle=0]{fig4.ps}
\caption
{Spectral region that includes the triplet lines
of Ne {\sc ix} and several lines of Fe {\sc xvii}. The
solid line is HD 98800; the dashed line is TW Hya.}
\end{figure}

\begin{figure}[htb]
\includegraphics[scale=.8,angle=0]{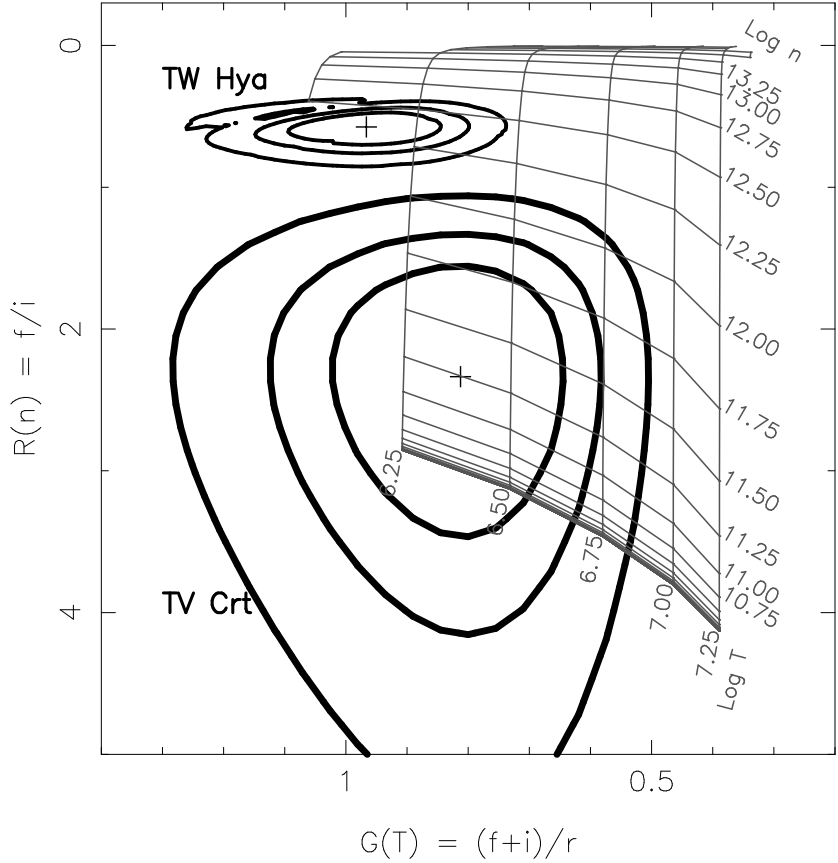}
\caption
{Line ratios $R=f/i$ vs.\ $G=(f+i)/r$ within the Ne {\sc ix} triplet, for
  HD 98800 = TV Crt (lower contours) and TW Hya (upper contours). The
  grid overlaid on the plot ($\log{n}$ vs.\ $\log{T}$), which is
  based on a density-dependent APED model (Brickhouse, pvt.\
  comm.), illustrates how the
  ratio $R$ serves as a diagnostic of the density of the gas at the
  temperature indicated by $G$. The
  central, middle, and outer contours
  represent 68\%, 90\%, and 99\% confidence levels in the
  measured values of $R$ and $G$. }
\end{figure}

\end{document}